\documentclass[aps,prl,twocolumn,superscriptaddress,longbibliography,floatfix]{revtex4-1}

\usepackage[T1]{fontenc}
\usepackage{newtxtext}
\usepackage{newtxmath}

\usepackage{amsmath}
\usepackage{mathtools}
\usepackage{amsfonts}
\usepackage{bm}

\usepackage{graphicx}
\usepackage{makecell}

\usepackage{color}
\usepackage{xcolor}
\usepackage[colorlinks,
            citecolor=blue,
            linkcolor=red,
            urlcolor=blue]{hyperref}

\begin{document}

\title{Evidence for deconfined $U(1)$ gauge theory at the transition between toric code and double semion}

\author{Maxime Dupont}
    \affiliation{Department of Physics, University of California, Berkeley, California 94720, USA}
    \affiliation{Materials Sciences Division, Lawrence Berkeley National Laboratory, Berkeley, California 94720, USA}

\author{Snir Gazit}
    \affiliation{Racah Institute of Physics and the Fritz Haber Center for Molecular Dynamics, The Hebrew University, Jerusalem 91904, Israel}

\author{Thomas Scaffidi}
    \affiliation{Department of Physics, University of Toronto, Toronto, Ontario, M5S 1A7, Canada}

\begin{abstract}
    Building on quantum Monte Carlo simulations, we study the phase diagram of a one-parameter Hamiltonian interpolating between trivial and topological Ising paramagnets in two dimensions, which are dual to the toric code and the double semion. We discover an intermediate phase with stripe order which spontaneously breaks the protecting Ising symmetry. Remarkably, we find evidence that this intervening phase is gapless due to the incommensurability of the stripe pattern and that it is dual to a $U(1)$ gauge theory exhibiting Cantor deconfinement.
\end{abstract}

\maketitle

\textit{Introduction.}--- At first sight, nontrivial bosonic symmetry-protected topological (SPT) phases~\cite{Chen1604,Chen2011,Chen2011b,YuanMing2012,PhysRevB.91.134404,pollmann2010,pollmann2012} look very similar to their trivial counterparts since they share the same symmetries, behave in the same way in the bulk, and do not possess a local order parameter. Several tools were proposed to unveil the differences between these phases, like comparing their edge properties, or looking at their entanglement spectrum or their many-body wave function  directly~\cite{Chen1604,Chen2011,Chen2011b,YuanMing2012,PhysRevB.91.134404,pollmann2010,pollmann2012,Cenke2014,ringel2015,PhysRevB.93.115105,PhysRevB.91.195134,PhysRevB.89.195122}. Another way is to gauge the protecting symmetry, since each SPT class is dual to a different Dijkgraaf-Witten gauge theory~\cite{dijkgraaf1990,Chen1604,Chen2011,Chen2011b,levin2012,PhysRevLett.114.031601}. As a result, if one attempts to interpolate from one class to the other, something drastic must happen on the way: either a quantum phase transition or an intermediate phase of matter which breaks spontaneously the protecting symmetry.

Exploring quantum phase transitions featuring SPTs is therefore a good place to look for exotic quantum criticality. In fact, transitions between different SPT phases~\cite{Grover2013,Lu_2014,morampudi2014,tsui2015,tsui2015b,you2016,he2016,you2018,tsui2017,geraedts2017,bi2019,bultinck2019,gozel2019,zeng2020}, and transitions between SPTs and symmetry-broken states~\cite{PhysRevB.83.174409,grover2012quantum,PhysRevB.91.235309,PhysRevLett.118.087201,parker2017,parker2018,parker2019,verresen2017,verresen2018,2019arXiv190506969V,PhysRevB.103.L100207}, have both attracted tremendous attention. Most of the existing work on transitions between SPTs has focused on continuous symmetries with ``large'' symmetry groups such as $\mathrm{O}(N)$, and relations with deconfined quantum criticality have been established in that context~\cite{Senthil1490,PhysRevX.7.031051,he2017,you2016,he2016,you2018,geraedts2017,bi2019,zeng2020}. On the other hand, the study of microscopic models with discrete symmetries has mostly been limited to one dimension ($1$D)~\cite{tsui2017,verresen2017}.

In this Letter, we investigate the quantum phase diagram of a one-parameter Hamiltonian interpolating between trivial and topological Ising ($\mathbb{Z}_2$) paramagnets in $2$D, which are dual~\cite{levin2012} to the toric code (TC)~\cite{KITAEV20032} and the double semion (DS)~\cite{PhysRevB.71.045110,PhysRevB.90.115129,PhysRevB.90.195148,PhysRevB.92.155105}, respectively. Unlike many other transitions between topological phases, this transition cannot be described in terms of anyon condensation~\cite{doi:10.1146/annurev-conmatphys-033117-054154}, and one has to resort to numerical studies~\cite{morampudi2014,huang2016,PhysRevB.97.195124,xu2018}. Although the double semion model itself has a sign problem that was proven to be irremediable~\cite{hastings2016,smith2020}, we have developed a sign-problem-free quantum Monte Carlo algorithm~\cite{dupont2020} which takes advantage of its SPT formulation. This allows us to access system sizes an order of magnitude larger than previous work, which relied on exact diagonalization~\cite{morampudi2014}. We find evidence for an intermediate incommensurate stripe phase which is dual to a deconfined $U(1)$ gauge theory, and which therefore evades Polyakov's result on the confinement of compact $U(1)$ gauges theories in $(2+1)$D~\cite{POLYAKOV1977429}. This is an observation of ``Cantor deconfinement'' in a microscopic system~\cite{PhysRevLett.64.92,fradkin2004,papanikolaou2007,schlittler2015,PhysRevLett.125.257204}.

\begin{figure}[!b]
    \center
    \includegraphics[width=1.0\columnwidth,clip]{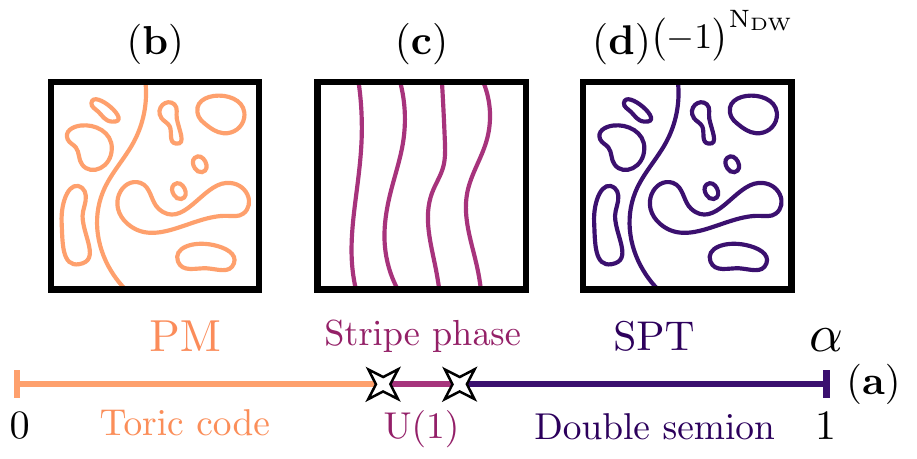}
    \caption{(\textbf{a}) Phase diagram of the model~\eqref{eq:ham} with intermediate stripe phase centered around $\alpha=1/2$. Domain walls, which separate up and down regions of $\sigma^z$, are represented for typical configurations, in the case of (\textbf{b}) the trivial paramagnetic phase (PM), (\textbf{c}) the stripe ordered phase, and (\textbf{d}) the topological paramagnetic phase (SPT).}
    \label{fig:model}
\end{figure}

\textit{Model.}--- A conventional $\mathbb{Z}_2$ paramagnet is described by the simple Hamiltonian, $\mathcal{H}_\mathrm{tr}=-\sum_j\sigma^x_j$, where $\sigma_j^{x,y,z}$ are Pauli matrices that live on the sites $j$ of the triangular lattice~\cite{supplemental}. It has a single gapped ground state $\left|{\psi_{\text{tr}}}\right\rangle$, which is an equal superposition of all $\sigma^z$ configurations. Since domain walls of Ising spins on the triangular lattice form closed nonintersecting domain walls on the dual honeycomb lattice, we can equally think of $\left|{\psi_{\text{tr}}}\right\rangle$ as an equal superposition of all domain wall configurations, see Fig.~\ref{fig:model}, which we denote symbolically as $\left|{\psi_{\text{tr}}}\right\rangle=\sum_{\text{dw}}\left|{\text{dw}}\right\rangle$~\cite{basis_domainwall}.

In $2$D, there exists a second type of $\mathbb{Z}_2$ paramagnet, which is fundamentally different from the trivial one as long as the $\mathbb{Z}_2$ symmetry is preserved~\cite{Chen1604,Chen2011,Chen2011b,levin2012}. A parent Hamiltonian for this topological phase is given by $\mathcal{H}_\mathrm{top}=\mathcal{U}^\dagger\mathcal{H}_\mathrm{triv}\mathcal{U}$, where $\mathcal{U}=(-1)^{N_\mathrm{dw}}$ is a unitary operator giving the parity of the number of domain walls $N_\mathrm{dw}$, see the Supplemental Material (SM) for an explicit form of $\mathcal{H}_\mathrm{top}$~\cite{supplemental}. This Hamiltonian also has a single gapped ground state which is a superposition of all domain wall configurations. The only difference with the trivial paramagnet is that each domain wall comes with a $-1$ fugacity, i.e., $\left|{\psi_{\text{top}}}\right\rangle=\sum_{\mathrm{dw}}(-1)^{N_\mathrm{dw}}\left|{\mathrm{dw}}\right\rangle$.

In this work, we interpolate between the two phases with the following one-parameter Hamiltonian,
\begin{equation}
    \mathcal{H}=\left(1-\alpha\right)\mathcal{H}_\mathrm{tr}+\alpha\mathcal{H}_\mathrm{top},\quad\alpha\in[0,1].
    \label{eq:ham}
\end{equation}

\begin{figure}[!t]
    \center
    \includegraphics[width=1.0\columnwidth,clip]{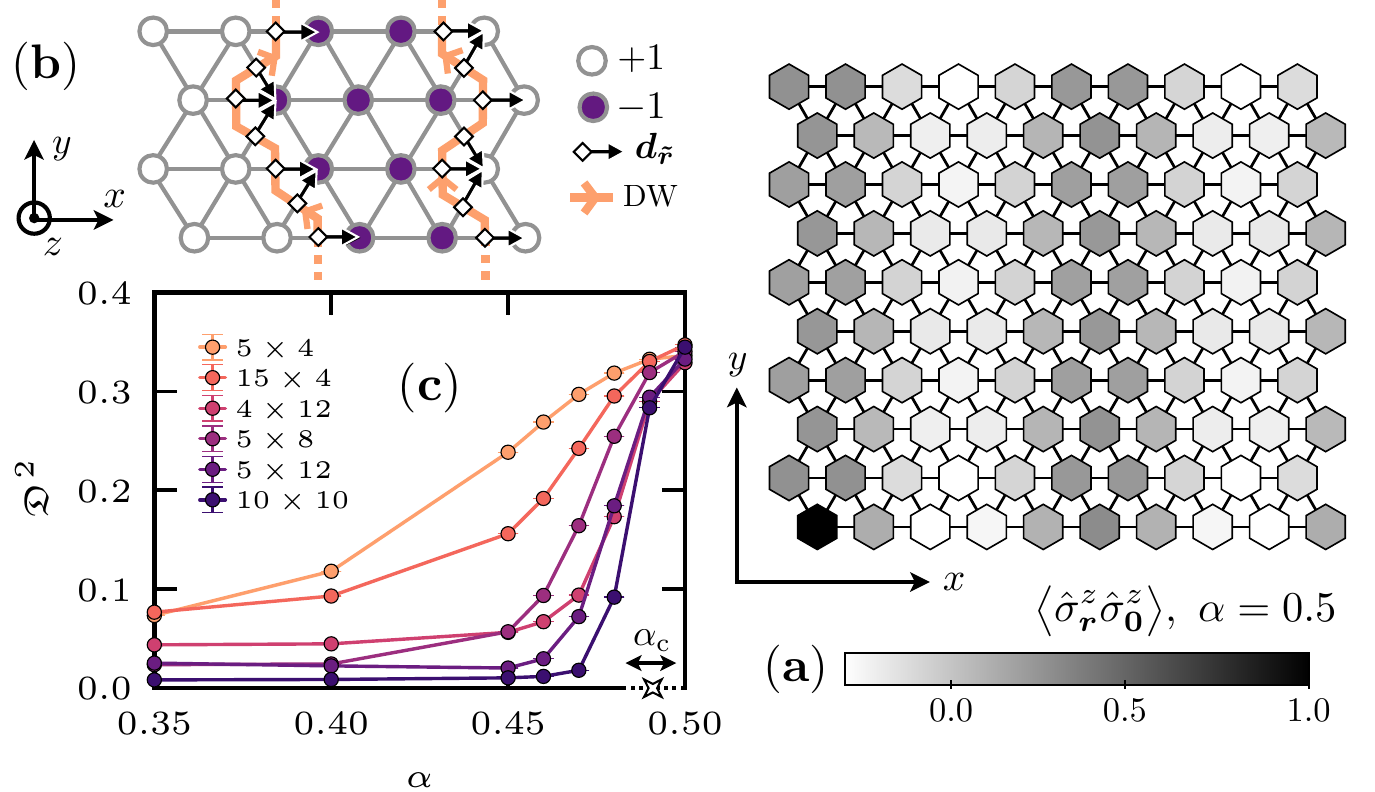}
    \caption{(\textbf{a}) Stripe structure revealed by $\langle\sigma^z_{\boldsymbol{r}}\sigma^z_{\boldsymbol{0}}\rangle$ at $\alpha=1/2$ for a periodic system of size $N=10\times{10}$. (\textbf{b}) Sketch of a spin configuration displaying two noncontractible oriented domain walls (orange lines). The black arrows are oriented unit length vectors $\boldsymbol{d}_{\boldsymbol{\tilde{r}}}$ orthogonal to the domain wall edges $\boldsymbol{\tilde{r}}$. (\textbf{c}) Square of the order parameter $\mathfrak{D}^2$ defined in Eq.~\eqref{eq:d2_order} versus $\alpha$ for increasing system sizes. The data is symmetric around $\alpha=1/2$ for $\alpha>1/2$.}
    \label{fig:order}
\end{figure}

\textit{Intermediate stripe order.}--- We investigate the model~\eqref{eq:ham} by means of quantum Monte Carlo simulations~\cite{supplemental,dupont2020}. Since $\mathcal{U}=\mathcal{U}^\dagger$, the phase diagram is symmetric around $\alpha=1/2$, see Fig.~\ref{fig:model}\,(a). It is therefore natural to start the analysis at $\alpha=1/2$, where the real-space two-point correlation $\langle\sigma^z_{\boldsymbol{r}}\sigma^z_{\boldsymbol{0}}\rangle$ reveals a stripe structure, as shown in Fig.~\ref{fig:order}\,(a) for a prototypical $N=10\times{10}$ system size.

We observe that the stripe pattern has a period given by $|\boldsymbol{Q}|\simeq{2\pi/5}$, where $\boldsymbol{Q}$ is the stripe wavevector (we take the lattice spacing equal to unity). However, the orientation of $\boldsymbol{Q}$ (given by the polar angle $\varphi$) and the precise value of the period vary depending on the finite-size system geometry~\cite{supplemental}. Depending on the system size, the stripe orientation belongs to one of two sets, which we call ``vertical'', with $\varphi=\mathbb{Z}2\pi/6$, and ``horizontal'', with $\varphi=(\mathbb{Z}+\frac{1}{2})2\pi/6$. Remarkably, these two sets of orientations are not related by symmetry, which is a good indication that the stripe order is only weakly pinned by the lattice. However, this makes a finite-size analysis based on peaks of the structure factor hazardous.

\begin{figure}[!t]
    \center
    \includegraphics[width=0.8\columnwidth,clip]{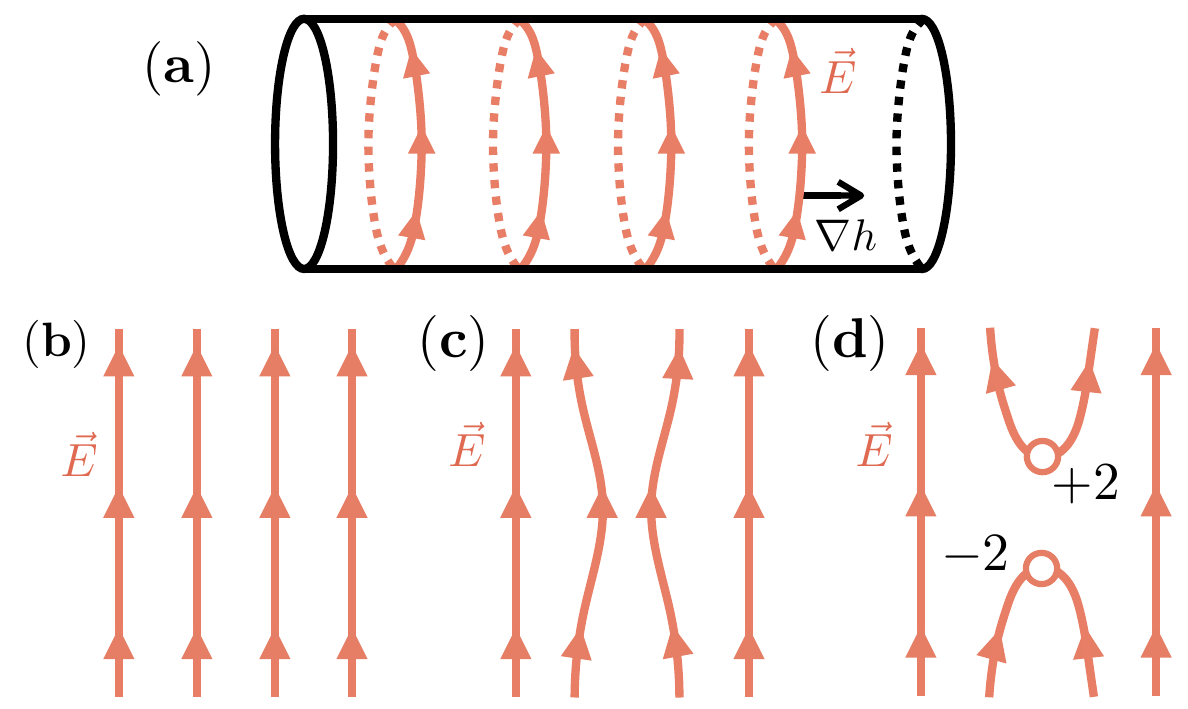}
    \caption{(\textbf{a}) The domain walls, once oriented, become electric field lines in the gauge description. The orange arrows indicate the direction of the electric field. The surgery process of going from (\textbf{b}) to (\textbf{d}) merges two noncontractible domain walls into one contractible one, and only becomes possible away from $\alpha=1/2$ (due to periodic boundary conditions, the top and bottom of each drawing should be identified). This process creates an electric dipole with charges $q=\pm{2}$.}
    \label{fig:efield}
\end{figure}

Instead, we define an order parameter which takes advantage of the domain wall representation. Whereas a paramagnetic phase has domain walls of all shapes and sizes, a perfect stripe phase only has noncontractible domain walls (NCDW) wrapping around the same handle of the torus, see Fig.~\ref{fig:model}\,(b,\,c,\,d). Let us define an order parameter $\mathfrak{D}$ that is proportional to the number of noncontractible domain walls $N_\mathrm{NCDW}$. In order to do this, it is useful to define an integer-valued height field $h$ living on the direct lattice which jumps by one unit every time a domain wall is crossed, and whose winding number around a given handle of the torus will give $N_\mathrm{NCDW}$. First, we give the same (arbitrary) orientation to each NCDW: For example, for vertical stripes, we choose an ``upwards'' orientation for each of them, see Figs.~\ref{fig:order}\,(b) and~\ref{fig:efield}\,(a)~\cite{supplemental}. This turns each domain wall strand into a vector which we call $\boldsymbol{E}_{\boldsymbol{\tilde{r}}}$ in anticipation of a gauge interpretation given later ($\boldsymbol{\tilde{r}}$ is an edge of the dual lattice). We can then define the vector field giving the gradient of $h$: $\nabla{h}\equiv\boldsymbol{d}_{\boldsymbol{\tilde{r}}}=\boldsymbol{E}_{\boldsymbol{\tilde{r}}}\times\boldsymbol{z}$, where $\boldsymbol{z}$ is the unit vector perpendicular to the plane.

A macroscopic number of NCDWs translates into a macroscopic tilt for the height field along the direction perpendicular to the domain walls. The squared norm of the tilt, defined as,
\begin{equation}
    \mathfrak{D}^2 = \left\langle\left(\frac{1}{N}\sum\nolimits_{\boldsymbol{\tilde{r}}}\boldsymbol{d}_{\boldsymbol{\tilde{r}}}\right)^2\right\rangle,
    \label{eq:d2_order}
\end{equation}
can thus be used as (the square of) our order parameter~\cite{order_parameter}. For a perfectly ordered stripe phase, there is a simple relation with the the stripe wave vector: $\mathfrak{D}^2=(3/2\pi)^2\boldsymbol{Q}^2$. As shown in Fig.~\ref{fig:order}\,(c) at $\alpha=1/2$, it takes a finite value $\mathfrak{D}^2\simeq{0.3}$ almost independently of the system size. Away from that point, our simulations do not allow us to draw a definite conclusion but from general arguments developed in the following, we expect a finite intermediate ordered phase centered around $\alpha=1/2$, albeit very small~\cite{supplemental}. In fact, the extrapolation of the data as $N\to+\infty$ is consistent with a jump of $\mathfrak{D}^2$ around $\alpha_\mathrm{c}\approx{0.48-0.49}$, suggestive of a first order transition between the stripe phase and the paramagnetic phase (and symmetrically at $\alpha_\mathrm{c}\approx{0.51-0.52}$ for the topological side).

\textit{Field theory.}--- Following previous works on stripe magnetism~\cite{PhysRevB.23.4615} and quantum dimer models~\cite{PhysRevB.69.224416,fradkin2004,Ardonne2004493,PhysRevB.65.024504,fradkin2013,moessner2011quantum}, we posit that the coarse-grained height field gives the phase of the local magnetization, $m^z(\boldsymbol{r})=|m^z|\cos(\pi h(\boldsymbol{r}))$, and that it is described by the Lagrangian,
\begin{equation}
    \begin{split}
        \mathcal{L} &= \frac{1}{2} \bigl(\partial_{\tau}h\bigr)^2 + V\bigl[h\bigr] + \lambda \cos\bigl(2\pi h\bigr),\\
        V\bigl[h\bigr] &= \frac{\rho_2}{2}\bigl(\nabla h\bigr)^2 + \frac{\rho_4}{2} \bigl(\nabla^2 h\bigr)^2 + \frac{g_4}{2}\bigl(\nabla h\bigr)^4 + \mathcal{L}_{6},
    \end{split}
    \label{eq:height_theory}
\end{equation}
where we have kept implicit the term which accounts for vortices of $h$. The stripe phase occurs for $\rho_2<0$, for which minimizing $V[h]$ leads to a tilt of the height field: $h(\boldsymbol{r},\tau)=\pi^{-1}\boldsymbol{Q}\cdot\boldsymbol{r}+\delta h(\boldsymbol{r},\tau)$, where $\delta{h}$ are the fluctuations around the perfectly tilted configuration. The orientation of $\boldsymbol{Q}$ is determined by the lowest order terms allowed on a triangular lattice: $\mathcal{L}_{6}=-g_6|\nabla{h}|^6\cos(6\varphi)-g_{12}|\nabla{h}|^{12}\cos(12\varphi)$, where $\varphi$ is the polar angle of $\nabla h$. Since we find both ``vertical'' and ``horizontal'' stripes for finite-size systems, we can conclude that $g_6$ is subdominant compared to $g_{12}$, leading to two different sets of six minima with almost degenerate values of $V[h]$. 

Two different scenarios are possible in order to melt the stripe order. The first one is to tune $\rho_2$ to zero, which continuously tunes $\boldsymbol{Q}\to\boldsymbol{0}$ until the multicritical Lifshitz (also known as Rokhsar-Kivelson) point  $\rho_2=0$~\cite{kivelson1988,PhysRevB.69.224416,fradkin2004,Ardonne2004493,PhysRevB.65.024504,fradkin2013,PhysRevB.83.125114,moessner2011quantum}. The second one is to fix $\rho_2$ but to increase the vortex fugacity, whose proliferation should mark the phase transition to a paramagnetic phase. We propose that this second scenario is the one at play at $\alpha_\mathrm{c}$.

Neglecting $\lambda$ for now, a long wavelength expansion around one of the minima of $V[h]$ leads to the following Goldstone theory,
\begin{equation}
    \mathcal{L} = \frac{1}{2} \bigl(\partial_{\tau}\delta h\bigr)^2 + \frac{v_{\mathrm{L}}^2}{2}\bigl(\partial_\mathrm{L} \delta h\bigr)^2 + \frac{v_{\mathrm{T}}^2}{2}\bigl(\partial_\mathrm{T} \delta h\bigr)^2 ,
    \label{eq:lowenergyTh}
\end{equation}
where $\mathrm{L}$ and $\mathrm{T}$ stand for the direction longitudinal and transverse to $\boldsymbol{Q}$, respectively~\cite{trans_stiffness}. The corresponding emergent continuous $U(1)$ symmetry is given by $h\rightarrow{h+c}$, with $c\in\mathbb{R}$ and describes the longitudinal translation of the stripe pattern with respect to the lattice.

We note that Eq.~\eqref{eq:height_theory} is dual to a compact $U(1)$ gauge theory for an electric field given by $\boldsymbol{E}=\boldsymbol{z}\times\nabla{h}$~\cite{PhysRevB.21.5212,doi:10.1142/S0217984990000295,PhysRevB.69.224416,fradkin2004,Ardonne2004493,PhysRevB.65.024504,fradkin2013}. Due to $\rho_2$ being negative, this theory has an unconventional ``Mexican hat'' electric energy density which goes like $-\boldsymbol{E}^2+\boldsymbol{E}^4$. This favors a finite density of electric field lines in the ground state, which correspond to the noncontractible domain walls of the stripe phase, see Fig.~\ref{fig:efield}\,(a).

An expansion around this configuration leads to the photon of Eq.~\eqref{eq:lowenergyTh}. A vertex operator of the type $\mathrm{e}^{i2\pi{ph}}$ maps to a monopole of charge $p$, and a vortex for $h$ of vorticity $q$ maps to a charge-$q$ electric charge ~\cite{fradkin2013}. We also provide in the SM a more microscopic justification for a $U(1)$ gauge theory description of the stripe phase~\cite{supplemental}.

We now turn our attention to $\lambda$, following Ref.~\onlinecite{fradkin2004}. This term imposes discrete values for the height field, and corresponds to the addition of $p=1$ monopoles in the gauge theory. When $\boldsymbol{Q}$ is incommensurate, $\lambda$ is irrelevant, and the gapless mode of Eq.~\eqref{eq:lowenergyTh} survives. This gapless regime is therefore dual to a deconfined $U(1)$ gauge theory, in which test charges experience logarithmic interactions. By contrast, when $\boldsymbol{Q}$ is commensurate, $\lambda$ is relevant and gaps out the photon, leading to a confined phase. As we will now show, we find good numerical evidence for a gapless Goldstone boson described by Eq.~\eqref{eq:lowenergyTh}, and we thus surmise that $\lambda$ is irrelevant, or very weakly relevant~\cite{small_gap}, leading to a deconfined $U(1)$ gauge theory.

\begin{figure}[!t]
    \center
    \includegraphics[width=1.0\columnwidth,clip]{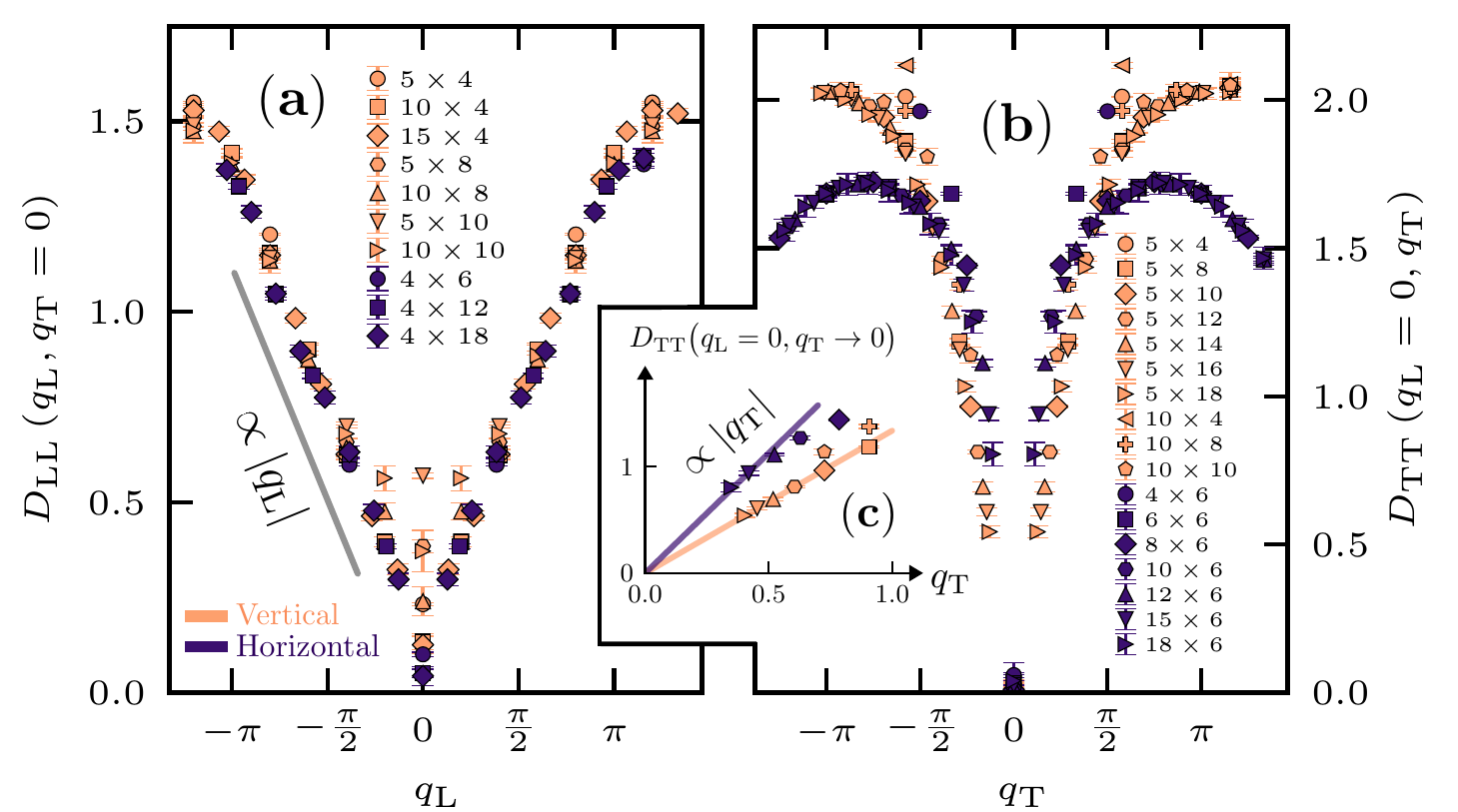}
    \caption{$D_{ab}(\boldsymbol{q})$ at $\alpha=1/2$ for various system sizes ($L_x\times{L_y}$) leading to vertical (orange) and horizontal (violet) stripes. (\textbf{a}) Longitudinal correlation $D_\mathrm{LL}(q_\mathrm{L},q_\mathrm{T}=0)\propto |q_\mathrm{L}|$ for $q_\mathrm{L}\to{0}$ and (\textbf{b}) transverse correlation $D_\mathrm{TT}(q_\mathrm{L}=0,q_\mathrm{T})\propto |q_\mathrm{T}|$ for $q_\mathrm{T}\to{0}$, in good agreement with Eq.~\eqref{eq:dd_corr_ft}. (\textbf{c}) Same data as (\textbf{b}) for the smallest $q_\mathrm{T}$ value for each system size, highlighting agreement with the linear prediction.}
    \label{fig:qspace_dd_corr_cuts}
\end{figure}

We probe the Goldstone boson by computing numerically the following two-point function:
\begin{equation}
    D_{ab}\bigl(\boldsymbol{q}\bigr)=\sum\nolimits_{\boldsymbol{\tilde{r}}}\mathrm{e}^{-i\boldsymbol{q}\cdot\boldsymbol{\tilde{r}}}\Bigl(\Bigl\langle d^a_{\boldsymbol{\tilde{r}}}d^b_{\boldsymbol{\tilde{0}}}\Bigr\rangle - \Bigl\langle d^a_{\boldsymbol{\tilde{r}}} \Bigr\rangle \Bigl\langle d^b_{\boldsymbol{\tilde{0}}}\Bigr\rangle \Bigr) ,
    \label{eq:dd_corr}
\end{equation}
which is displayed in Fig.~\ref{fig:qspace_dd_corr_cuts} at $\alpha=1/2$, see also SM~\cite{supplemental}. Identifying $\boldsymbol{d}_{\boldsymbol{\tilde{r}}}$ with the height gradient $\nabla{h}$, one can derive from Eq.~\eqref{eq:lowenergyTh} the following prediction~\cite{supplemental}:
\begin{equation}
    D_{ab}\bigl(\boldsymbol{q}\bigr)=\frac{q_a q_b}{2\sqrt{(v_{\mathrm{L}}q_{\mathrm{L}})^2 + (v_{\mathrm{T}} q_{\mathrm{T}})^2}}.
    \label{eq:dd_corr_ft}
\end{equation}
Note that, in the gauge picture, this is nothing but an electric field correlator~\cite{PhysRevLett.91.167004}. The linear behavior $D_\mathrm{aa}(q_\mathrm{a},q_\mathrm{b}=0)\sim q_a/2v_a$ for $q_a\to{0}$ is observed in Fig.~\ref{fig:qspace_dd_corr_cuts}, and we estimate the velocities to be $v_\mathrm{L}\simeq{1.2}$ and $v_\mathrm{T}\simeq{0.3}$. Based on this, we extract the following field theory parameters~\eqref{eq:height_theory}: $g_6\simeq{0}$, $g_{12}\simeq{6.5}$, $\rho_2\simeq-{0.72}$, and $g_4\simeq{2.25}$~\cite{supplemental}.

\textit{Melting the stripe order.---} Now that we have a good picture of the stripe order at $\alpha=1/2$, we can ask how this order gives way to a trivial (respectively topological) paramagnet, when the parameter $\delta\alpha\equiv{1/2}-\alpha$ is tuned to positive (respectively negative) values. Terms proportional to $\delta\alpha$ are the only ones allowing spin flips that change the number of domain walls by $\pm{1}$~\cite{supplemental}. Therefore, at $\delta\alpha=0$, the number of noncontractible domain walls is almost~\cite{process_ndw} conserved (see numerical evidence in the SM~\cite{supplemental}), and the physics is simply described by their vibrations with Eq.~\eqref{eq:lowenergyTh}.

By contrast, for $\delta\alpha\neq{0}$, it becomes possible to create contractible domain walls, either from the vacuum or by doing surgery on two noncontractible domain walls, see Fig.~\ref{fig:efield}\,(b,\,c,\,d). As $|\delta\alpha|$ is increased, these contractible domain walls will proliferate, eventually leading to a condensate of domain walls of all shapes and sizes, i.e., a paramagnetic phase. The negative sign of $\delta\alpha$ on the topological side ensures that each contractible domain wall occurs with the appropriate $-1$ factor.

As seen in Fig.~\ref{fig:efield}\,(d) (see also SM~\cite{supplemental}), in the gauge picture, a contractible domain wall can be described as a $\pm{2}$ electric dipole created to screen the background electric field~\cite{pair_creation}. This means that $\delta\alpha$ controls the fugacity of $\pm{2}$ electric dipoles and the transition to the paramagnetic phases occurs when these charges condense, giving rise to a ``Higgs'' phase. Since the condensation of charge $q$ particles in a $U(1)$ gauge theory leads to a $\mathbb{Z}_q$ gauge theory~\cite{PhysRevD.19.3682,sachdev1991,PhysRevB.49.5200,deWild}, we recover that the paramagnetic phases are dual to $\mathbb{Z}_2$ gauge theories, as expected. The condensation of charge-$2$ matter in a $(2+1)$D compact $U(1)$ gauge theory was studied before in the case of $\rho_2>0$, in which case the $U(1)$ gauge theory is confined due to monopoles~\cite{PhysRevD.19.3682,PhysRevLett.89.277004,PhysRevB.66.205104,PhysRevB.67.205104,Ardonne2004493}. However, to the best of our knowledge, the nature of this transition in the case of $\rho_2<0$ has not been considered before, and is left for future work.

To sum up, transitioning from the trivial to the topological paramagnet requires changing the fugacity of contractible domain walls from $+1$ to $-1$~\cite{fugacity_sign}. In order to do so, the fugacity has to go through zero in the middle, leading to an intermediate stripe phase with (almost) no contractible domain walls. In the gauge picture, contractible domain walls are described by $\pm{2}$ electric dipoles, and the intermediate phase is thus described by a Coulomb phase with (almost) no electric charges.

Note that $q=\pm{2}$ are the smallest dynamical charges allowed by the Hamiltonian since $q=\pm{1}$ charges would require a dangling domain wall configuration (called $\pi$-flux excitation), which are only allowed as static excitations in the TC/DS models. In fact, the $q=\pm{1}$ charge survives as one of the gapped quasiparticles of the Higgs phases: It becomes the bosonic $e$ excitation of TC, and one of the semions of DS~\cite{levin2012}.

The other excitation to survive in the Higgs phases is the $p=1/2$ monopole, which is dual to $\cos(\pi{h})$ in the height language, and is created by $\sigma^z_j$ in the original microscopic model. The fact that such a fractional monopole is allowed can be traced back to the nontrivial mapping between Ising spins and domain walls: Translating all domain walls by one inter-domain-wall separation is a good symmetry for the domain walls ($h\rightarrow{h+1}$), but not for the spins (since up and down regions are interchanged in the process). The $p=1/2$ monopole is denoted $m$ and is bosonic in both toric code and double semion~\cite{levin2012}.

\textit{Discussion.}--- Naively, one might have expected an intermediate phase which breaks a discrete symmetry to be gapped, and dual to a confined theory. This would have been the case for a ferromagnetic phase for example, for which confinement is a natural consequence of the fact that only short domain walls exist. We have found instead a phase which breaks the Ising symmetry but that has nevertheless long, fluctuating domain walls which allow for a dual deconfined theory.

The nature of the transition between the stripe phase and the paramagnetic phases would require further work to be pinned down. Our current results point towards a first-order transition, but another possibility would be an intervening nematic phase, which breaks rotation but not translation symmetry. Besides, the behavior of entanglement entropy across the transitions could have unique properties, since both toric code and double semion have the same topological contribution, but the intermediate $U(1)$ gauge theory should have a logarithmic contribution instead~\cite{entanglement}. We also expect the stripe phase to have anomalous edge properties on the topological side, inspired by previous work on gapless SPTs~\cite{PhysRevLett.118.087201,parker2017,parker2018,parker2019,2019arXiv190506969V}.

Finally, our sign-problem-free Monte Carlo algorithm enables us to add a variety of other terms in the Hamiltonian and to study other classes of SPT protected by discrete symmetries~\cite{dupont2020}. This could enable us to tune $\rho_2$ towards the quantum Lifshitz point, and to study the ``Devil's staircase'' of commensurate-incommensurate transitions predicted to happen on the way~\cite{fradkin2004,papanikolaou2007}. Another potentially nearby multicritical point could be $\mathrm{QED}_3$ with $N_f=2$~\cite{PhysRevX.9.021022,PhysRevD.100.054514}, which was predicted to describe the transition between toric code and double semion in the presence of $\mathrm{SU}(2)$ symmetry~\cite{2013arXiv1307.8194B}. The closely related deconfined quantum critical point of the $J$-$Q$ model~\cite{PhysRevLett.98.227202} was in fact recently shown to appear at the tip of a helical valence bond phase which resembles the stripe phase presented in this work~\cite{PhysRevLett.125.257204}.

\begin{acknowledgments}
    \textit{Acknowledgments.}--- We are grateful to F. Alet, N. Bultinck, X. Cao, S. Capponi, E. Fradkin, B. Kang, A. Paramekanti, D. Poilblanc, F. Pollmann, P. Pujol, A. W. Sandvik, R. Vasseur, C. Xu and L. Zou for interesting discussions. M. D. was supported by the U.S. Department of Energy, Office of Science, Office of Basic Energy Sciences, Materials Sciences and Engineering Division under Contract No. DE-AC02-05-CH11231 through the Scientific Discovery through Advanced Computing (SciDAC) program (KC23DAC Topological and Correlated Matter via Tensor Networks and Quantum Monte Carlo). S. G. acknowledges support from the Israel Science Foundation, Grant No. 1686/18. T. S. acknowledges the financial support of the Natural Sciences and Engineering Research Council of Canada (NSERC), in particular the Discovery Grant [RGPIN-2020-05842], the Accelerator Supplement [RGPAS-2020-00060], and the Discovery Launch Supplement [DGECR-2020-00222]. This research used the Lawrencium computational cluster resource provided by the IT Division at the Lawrence Berkeley National Laboratory (Supported by the Director, Office of Science, Office of Basic Energy Sciences, of the U.S. Department of Energy under Contract No. DE-AC02-05CH11231). This research also used resources of the National Energy Research Scientific Computing Center (NERSC), a U.S. Department of Energy Office of Science User Facility operated under Contract No. DE-AC02-05CH11231. T.S. contributed to this work prior to joining Amazon.
\end{acknowledgments}

\bibliography{bibliography}

\begin{thebibliography}{101}
\bibitem [{bas()}]{basis_domainwall}%
  \BibitemOpen
  \href@noop {} {}\bibinfo {note} {Strictly speaking, since the mapping from
  spins to domain walls is $2$ to $1$, there should be an extra sum over the
  two possible spin orientations for a given domain wall
  configuration.}\BibitemShut {Stop}%
\bibitem [{ord()}]{order_parameter}%
  \BibitemOpen
  \href@noop {} {}\bibinfo {note} {A similar order parameter, called winding
  number, was used for a stripe phase on the triangular lattice in
  Ref.~\onlinecite{PhysRevLett.116.197201}.}\BibitemShut {Stop}%
\bibitem [{tra()}]{trans_stiffness}%
  \BibitemOpen
  \href@noop {} {}\bibinfo {note} {Note that the transverse stiffness term
  $(\partial_\mathrm{T}{\delta}{h})^2$ only arises due to $\mathcal{L}_6$, and
  the transverse velocity is therefore parametrically different from the
  longitudinal one~\cite{PhysRevLett.103.010404}.}\BibitemShut {Stop}%
\bibitem [{sma()}]{small_gap}%
  \BibitemOpen
  \href@noop {} {}\bibinfo {note} {A possible scenario would be a commensurate
  order with a very small gap, since it was shown that the gap is exponentially
  small in the weak-tilt regime~\cite{fradkin2004}.}\BibitemShut {Stop}%
\bibitem [{pro()}]{process_ndw}%
  \BibitemOpen
  \href@noop {} {}\bibinfo {note} {The only process which can change the number
  of domain walls is a rare event in which three domain walls come close and
  combine into one.}\BibitemShut {Stop}%
\bibitem [{pai()}]{pair_creation}%
  \BibitemOpen
  \href@noop {} {}\bibinfo {note} {This process of pair creation can be seen as
  a condensed matter version of the Schwinger
  effect~\cite{PhysRev.82.664}.}\BibitemShut {Stop}%
\bibitem [{fug()}]{fugacity_sign}%
  \BibitemOpen
  \href@noop {} {}\bibinfo {note} {It is not necessary to change the sign of
  the fugacity of non-contractible domain walls, since the parity of
  $N_\mathrm{NCDW}$ is always fixed in a given sector.}\BibitemShut {Stop}%
\end{thebibliography}

\appendix
\onecolumngrid
\setcounter{secnumdepth}{3}
\setcounter{figure}{0}
\setcounter{equation}{0}
\renewcommand\thefigure{S\arabic{figure}}
\renewcommand\theequation{S\arabic{equation}}
\newpage\clearpage

\setlength{\belowcaptionskip}{0pt}

\begin{center}
    \large\textbf{Supplemental material to ``\textit{Evidence for deconfined $U(1)$ gauge theory\\ at the transition between toric code and double semion}''}
\end{center}

\addvspace{5mm}
\begin{center}
    \begin{minipage}{0.8\textwidth}
        In this supplemental material, we provide the explicit form for the topological Hamiltonian, information regarding the quantum Monte Carlo simulations, an intuitive justification for the apparition of a stripe order, details about the prescription used to orient contractible domain walls, additional numerical data as well as additional details regarding the field theory. The numerical data include evidence for the incommensurability of the stripe order, the extrapolation of the order parameter $\mathfrak{D}^2$ in the thermodynamic limit (as $N\to+\infty$), an additional comparison between the microscopic model and the field theory predictions and a discussion regarding the average number of noncontractible domain walls at $\alpha=1/2$. The field theory section includes a part regarding the tilted phase, a dictionary translating between height and gauge theories, and a microscopic justification for a $U(1)$ gauge theory description of the stripe phase.
    \end{minipage}
\end{center}

\section{Explicit form of $\mathcal{H}_\mathrm{top}$}

\begin{figure*}[!h]
    \center
    \includegraphics[width=0.5\columnwidth,clip]{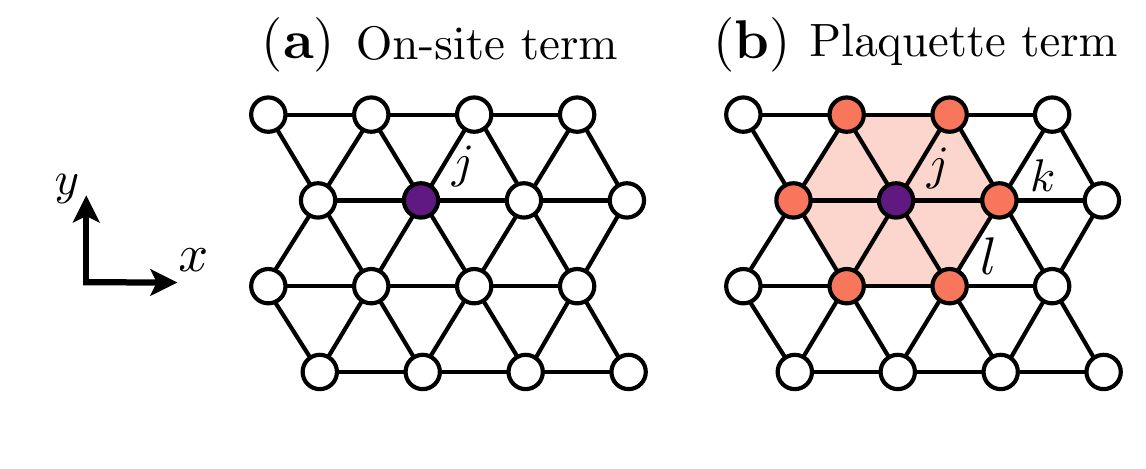}
    \caption{The one-parameter model of Eq.~(1) in the main text is defined on the triangular lattice and made of two distinct terms. (\textbf{a}) The first one describes a trivial Ising paramagnet with on-site terms and the second (\textbf{b}) describes a topological Ising paramagnet protected by the $\mathbb{Z}_2$ Ising spin-flip symmetry, with plaquette terms involving the six nearest neighbors of a given lattice site $j$, see Eq.~\eqref{eq:ham_top}.}
    \label{fig:lattice_model}
\end{figure*}

We give an explicit form for the topological Ising paramagnet Hamiltonian $\mathcal{H}_\mathrm{top}$, part of the one-parameter model of Eq.~(1) in the main text, interpolating between trivial and topological phases, and which is studied in this work. Following~\cite{Chen1604,Chen2011,Chen2011b,levin2012},
\begin{equation}
    \mathcal{H}_\mathrm{top}=-\sum\nolimits_{j}\Biggl(\sigma^x_j\prod\nolimits_{\bigtriangleup_{jkl}}i^{\frac{1}{2}\left(-1-\sigma^z_k\sigma^z_l\right)}\Biggr),
    \label{eq:ham_top}
\end{equation}
where the product runs over the six triangles containing the site $j$, see Fig.~\ref{fig:lattice_model}\,(b). This product has a simple interpretation in terms of domain walls: It gives a minus sign if flipping the spin of $\sigma_j$ changes the parity of the number of domain walls $N_\mathrm{DW}$.

\section{Quantum Monte Carlo simulations}

In the $\sigma^z$ basis, the model of Eq.~1 in the main text is only sign-problem-free for $\alpha\leq 1/2$. However, for periodic boundary conditions, its spectrum is exactly symmetric around $\alpha=1/2$ because of the unitary transformation $\mathcal{U}$ relating the trivial and topological terms~\cite{levin2012}. Therefore, as far as thermodynamic properties are concerned, the phase diagram is symmetric around $\alpha=1/2$ and we can restrict the numerical study to $\alpha\leq{1/2}$. It might seem surprising at first that we were able to find a sign-problem-free algorithm for a model that is dual to the double semion, given the existing work on the sign problem in this model~\cite{hastings2016,smith2020}. The reason why this was possible is that (i) we use periodic boundary conditions, and (i) $\pi$-flux excitations only appear as gapped, static excitations in the gauged models. Relaxing any of these conditions would most likely generate a sign problem. We use both projective~\cite{becca2017} and stochastic series expansion~\cite{sandvik2010,sandvik2019} quantum Monte Carlo to simulate the model. Details are available in Ref.~\onlinecite{dupont2020}.

\section{Intuition behind the stripe order}

In this section we provide an intuitive explanation for the apparition of the stripe order. The Hamiltonian (1) of the main text is simply a transverse field model for which the value of the transverse field on a given site can take two values, depending on whether flipping that spin changes the parity of $N_\mathrm{dw}$. The field amplitude is $1-2\alpha$ if it changes the $N_\mathrm{dw}$-parity, and $1$ if it does not. At $\alpha=1/2$, $N_\mathrm{dw}$-parity-changing spin flips become thus strictly disallowed, creating a local kinetic constraint which bears some similarities with quantum dimer~\cite{moessner2011quantum} and loop models~\cite{PhysRevResearch.2.033051}. Specifically, one cannot perform a spin flip that would create or annihilate a single domain wall, or that would merge two domain walls into one. The tendency of the ground state to maximize the number of flippable spins creates a nontrivial effective interaction for the domain walls: It is attractive at long distance (since a large ferromagnetic region has no flippable spins), and repulsive at short distance (since merging two domain walls is prohibited). In the spin language, this corresponds to an interaction that is ferromagnetic at short distance and antiferromagnetic at long distance, like in the anisotropic next-nearest neighbor Ising model (which is a textbook model for incommensurate stripe order)~\cite{SELKE1988213}.

\section{Orientation of contractible domain walls when computing $D_{ab}(\mathbf{q})$}

As explained in the main text, it is necessary to have a consistent prescription on how to orient domain walls in order to map them to a height configuration. For noncontractible domain walls (NCDWs), the situation is easy. We give them all the same (arbitrary) orientation: For example, for vertical stripes, we choose an ``upwards'' orientation for each of them, see Fig.~3 of the main text. In order to calculate the order parameter $\mathfrak{D}$, this is all we need to know, since contractible domain walls do not contribute to $\mathfrak{D}$.

However, in order to compute height gradient correlators $D_{ab}(\boldsymbol{q})$ (see Eq.~(5) of the main text), it is also necessary to orient contractible domain walls. As explained in the main text and in Sec.~\ref{ExplicitMapping}, a proper accounting of the orientation of contractible domain walls would require the introduction of electric dipoles in the gauge picture, or equivalently of vortices in the height field. However, there are very few contractible domain walls at $\alpha=1/2$ anyway (see numerical results below), so for simplicity we decided to give the same arbitrary orientation (namely counterclockwise) to all contractible domain walls when computing $D_{ab}(\boldsymbol{q})$ numerically. We do not expect this choice to have a substantial effect on the numerical results.

\section{Additional numerical data}

\subsection{Structure factor}

\begin{figure*}[!t]
    \center
    \includegraphics[width=0.8\columnwidth,clip]{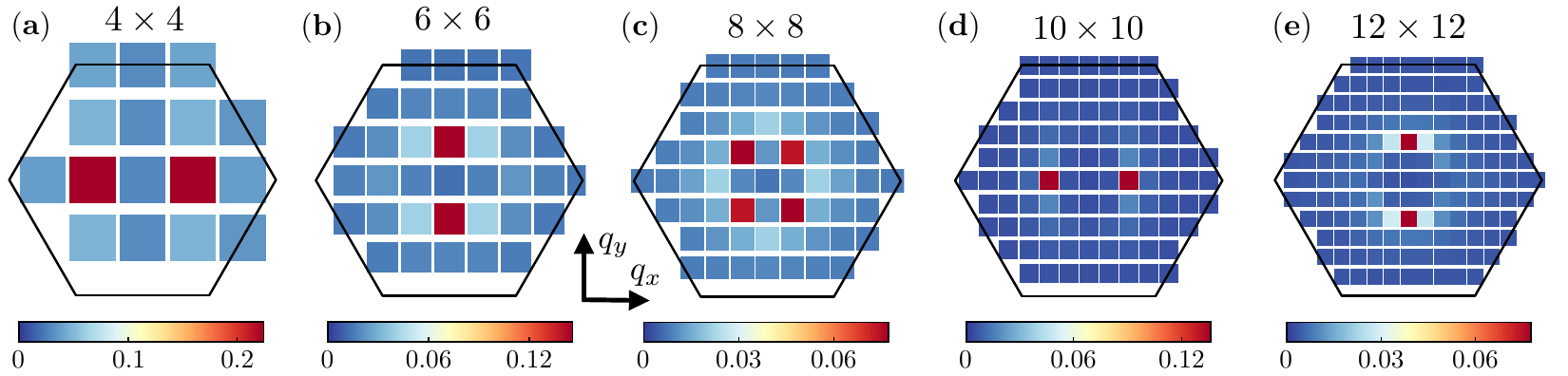}
    \caption{Structure factor based on the real-space correlation function $\langle\sigma^z_{\boldsymbol{r}}\sigma^z_{\boldsymbol{0}}\rangle$, computed from Eq.~\eqref{eq:corr_zz_q} at $\alpha=1/2$ for system sizes (\textbf{a}) $N=4\times 4$, (\textbf{b}) $N=6\times 6$, (\textbf{c}) $N=8\times 8$, (\textbf{d}) $N=10\times 10$, and (\textbf{e}) $N=12\times 12$. Going from one system size to the next, the position in the Brillouin zone of the maximum intensity changes, but remains around $|\boldsymbol{Q}|\simeq 2\pi/5$. Precisely, (\textbf{a}) $\boldsymbol{Q}=(\pm\pi/2, 0)$, (\textbf{b}) $\boldsymbol{Q}=(\pm 2\pi/3\sqrt{3}, 0)$, (\textbf{c}) $\boldsymbol{Q}=(\pm \pi/4,\pm \pi/2\sqrt{3})$, (\textbf{d}) $\boldsymbol{Q}=(\pm 2\pi/5,0)$, and (\textbf{e}) $\boldsymbol{Q}=(\pm 2\pi/3\sqrt{3}, 0)$. Following the notation explained in the text, these system sizes display ``vertical'' stripes for (\textbf{a}),(\textbf{d}) and ``horizontal'' stripes for (\textbf{b}),(\textbf{c}),(\textbf{e}). }
    \label{fig:zq_sizes}
\end{figure*}

To probe long-range order, one can look at the structure factor, corresponding to the Fourier transform of the two-point correlation function $\langle\sigma^z_{\boldsymbol{r}}\sigma^z_{\boldsymbol{0}}\rangle$,
\begin{equation}
    S\bigl(\boldsymbol{q}\bigr)=\frac{1}{N}\sum\nolimits_{\boldsymbol{r}}\mathrm{e}^{-i\boldsymbol{q}\cdot\boldsymbol{r}}\Bigl\langle\sigma^z_{\boldsymbol{r}}\sigma^z_{\boldsymbol{0}}\Bigr\rangle.
    \label{eq:corr_zz_q}
\end{equation}
It is displayed in Fig.~\ref{fig:zq_sizes} for various system sizes at $\alpha=1/2$. Going from one system size to the next, the position in the Brillouin zone of the maximum intensity changes, although it always remains at $|\boldsymbol{Q}|\approx 2\pi/5$. This makes it difficult to do any consistent finite-size study of what appears to be a putative incommensurate ordered phase.

Denoting the polar angle of $\boldsymbol{Q}$ by $ \varphi$, we find ``vertical'' stripes (with $\varphi=\mathbb{Z}2\pi/6$) when $L_x$ is close to a multiple of $5$, and ``horizontal'' stripes (with $\varphi=(\mathbb{Z}+\frac{1}{2})2\pi/6$) when $L_y$ is a multiple of $6$. Remarkably, these two sets of orientations are not related by symmetry.

\subsection{Thermodynamic extrapolation of $\mathfrak{D}^2$}

\begin{figure}[!t]
    \center
    \includegraphics[width=0.8\columnwidth,clip]{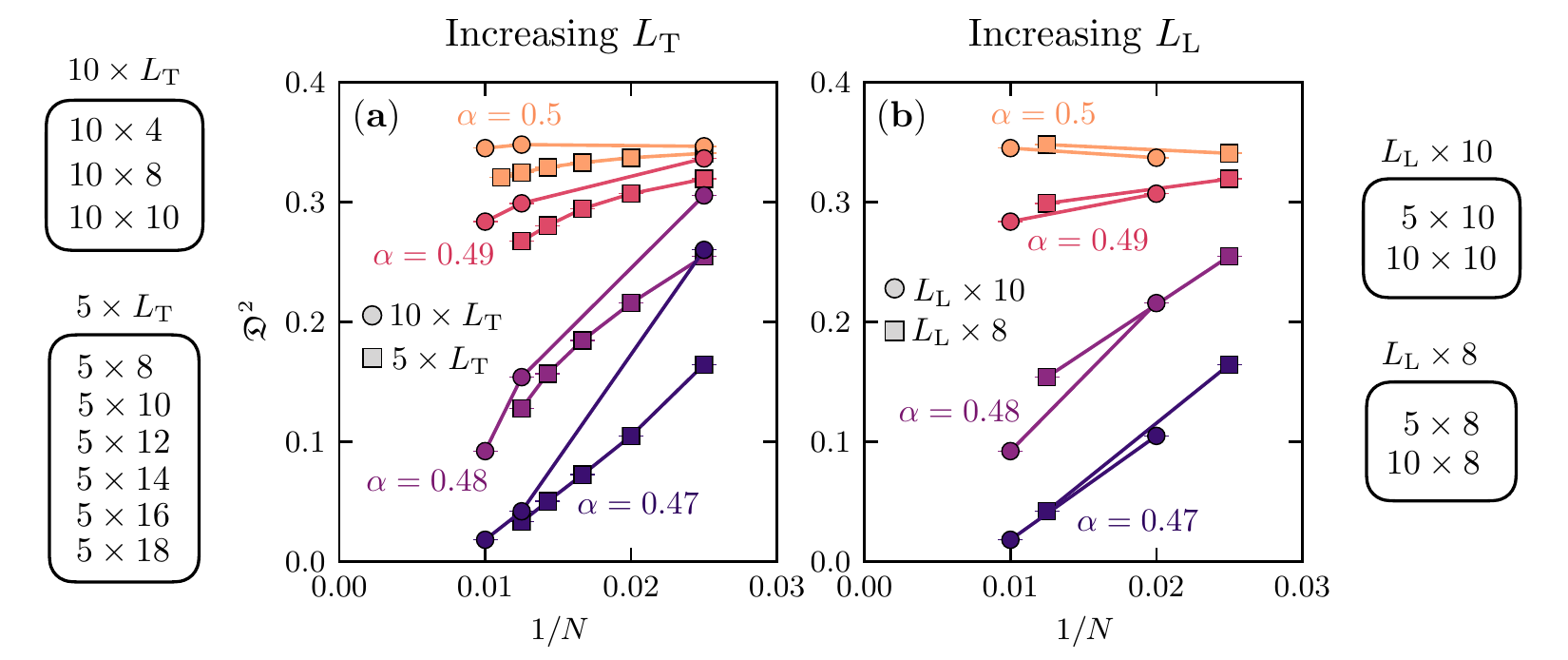}
    \caption{Order parameter value $\mathfrak{D}^2$ versus the inverse system size $1/N$ for different values of $\alpha$. (\textbf{a}) Two different series of system sizes $N=10\times L_\mathrm{T}$ and $N=5\times L_\mathrm{T}$, with increasing length $L_\mathrm{T}$ along the stripes direction. (\textbf{b}) Two different series of system sizes $N=L_\mathrm{L}\times 10$ and $N=L_\mathrm{L}\times 8$ with increasing length $L_\mathrm{L}$ transverse to the stripes direction. The order parameter takes a finite value $\mathfrak{D}^2\simeq 0.3$ at $\alpha=0.5$ and goes to zero as $N\to+\infty$ at $\alpha=0.47$, but it is difficult to draw any definite conclusion for intermediate values of $\alpha$.}
    \label{fig:order_vs_N}
\end{figure}

Additionally to Fig.~2\,(c) of the main text showing the order parameter value $\mathfrak{D}^2$ versus $\alpha$ for different system sizes, we display in Fig.~\ref{fig:order_vs_N} the order parameter $\mathfrak{D}^2$ versus $1/N$ for different values of $\alpha$ close to $\alpha=1/2$. In particular, we consider two cases corresponding to increasing one length of the system along or perpendicular to the stripes. In each case, we show that $\mathfrak{D}^2\simeq 0.3$ as $N\to+\infty$ at $\alpha=0.5$. We also show that it goes to zero as $N\to+\infty$ for $\alpha=0.47$. However, it is difficult to draw any definite conclusion for intermediate values of $\alpha$ except that the stripe ordered phase is relatively small. The data is consistent with a jump of $\mathfrak{D}^2$ around $\alpha_\mathrm{c}\approx 0.48-0.49$, indicative of a first order transition, as discussed in the main text.

\subsection{Field theory comparison for $D_{ab}(\mathbf{q})$}

\begin{figure}[!t]
    \center
    \includegraphics[width=0.8\columnwidth,clip]{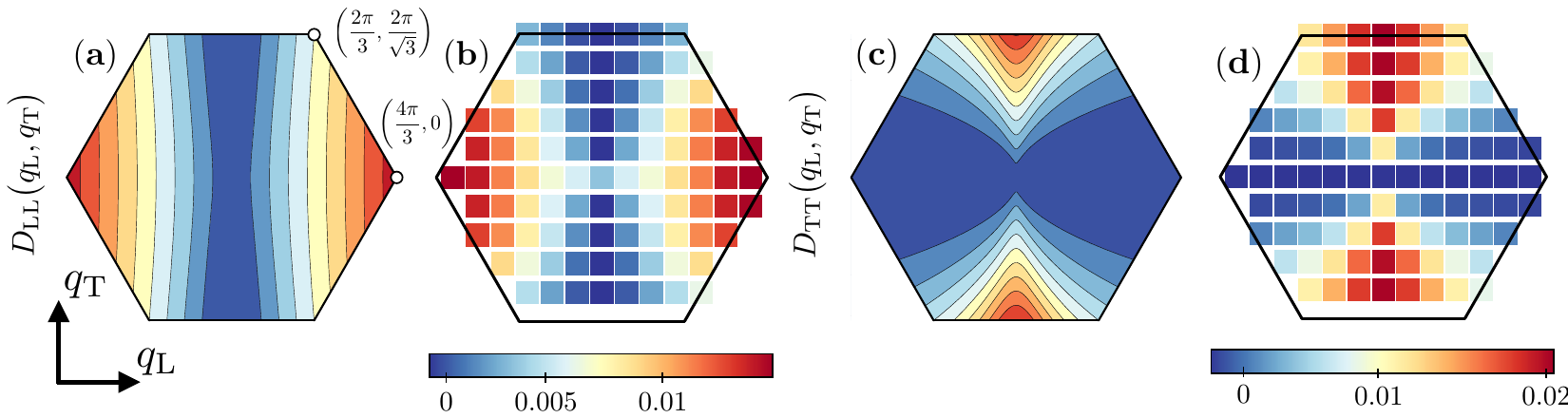}
    \caption{(\textbf{a},\,\textbf{c}) Color map intensity of the field theory prediction for the correlation function of Eq.~(6) in the main text, using $v_\mathrm{L}=1.2$ and $v_\mathrm{T}=0.3$. (\textbf{b},\,\textbf{d}) Color map intensity of the correlation function of Eq.~(5) in the main text computed in quantum Monte Carlo for a system of size $10\times 10$ at $\alpha=1/2$ There is a good qualitative agreement between the field theory predictions and the data of the microscopic model.}
    \label{fig:qspace_dd_corr_cmap}
\end{figure}

Additionally to Fig.~4 of the main text, aiming at comparing the microscopic model with the field theory predictions, we display in Fig.~\ref{fig:qspace_dd_corr_cmap} the full color map intensity of the correlation functions $D_\mathrm{LL}$ and $D_\mathrm{TT}$ (see Eqs. (5) and (6) of the main text). There is a good qualitative agreement, and we have used $v_\mathrm{L}=1.2$ and $v_\mathrm{T}=0.3$ for the field theory prediction, as estimated from Fig.~4 in the main text.

\subsection{Number of noncontractible domain walls at $\alpha=1/2$}

\begin{figure}[!t]
    \center
    \includegraphics[width=0.6\columnwidth,clip]{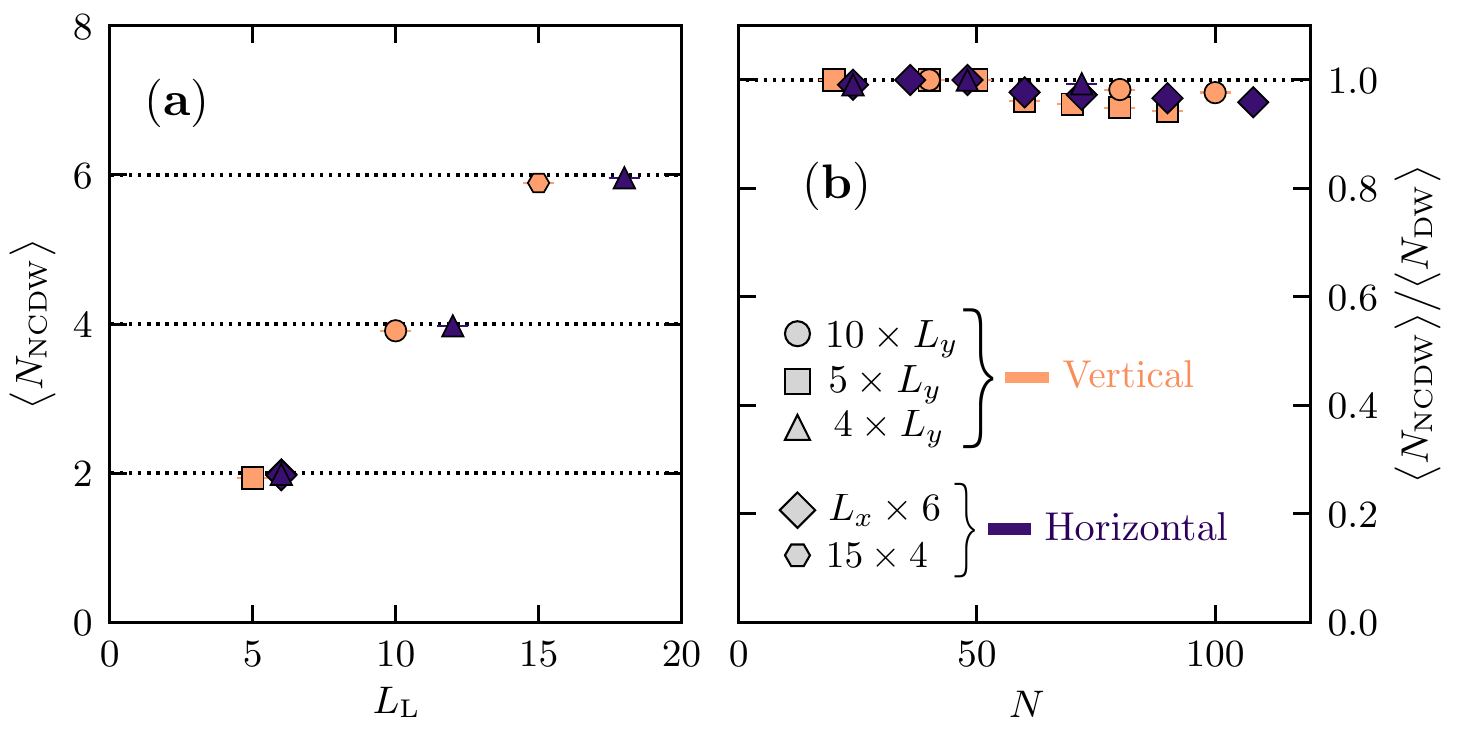}
    \caption{(\textbf{a}) Expectation value of the number of noncontractible domain walls $\langle N_\mathrm{NCDW}\rangle$ for various system sizes at $\alpha=1/2$. It is displayed against the length $L_\mathrm{L}$ of the system that is perpendicular to the orientation of the stripes. This number is extremely close to an integer, meaning that it is almost conserved. (\textbf{b}) Ratio of the number of noncontractible domain walls over the total number of domain walls (contractible and noncontractible together) $\langle N_\mathrm{NCDW}\rangle/\langle N_\mathrm{DW}\rangle$ for various system sizes at $\alpha=1/2$. This number is extremely close to $1$, which means that the system mostly consists of noncontractible domain walls at $\alpha=1/2$, supporting the microscopic origin of the stripe order of noncontractible domain walls wrapping around the torus. The orange symbols correspond to system sizes with vertical stripes while the violet symbols correspond to those with horizontal stripes.}
    \label{fig:ndw_vs_N}
\end{figure}

In the quantum Monte Carlo simulations, we compute the expectation value of the number of contractible and noncontractible domain walls: $\langle N_\mathrm{DW}\rangle$ and $\langle N_\mathrm{NCDW}\rangle$, respectively. It is shown in Fig.~\ref{fig:ndw_vs_N}\,(a) at $\alpha=1/2$ for various system sizes. We find that the average number of noncontractible domain walls is extremely close to an integer, meaning that it is almost conserved. Moreover the ratio of the number of noncontractible domain walls over the total number of domain walls $\langle N_\mathrm{NCDW}\rangle/\langle N_\mathrm{DW}\rangle$ is extremely close to $1$, as displayed in Fig.~\ref{fig:ndw_vs_N}\,(b). This means that the system mostly consists of noncontractible domain walls at $\alpha=1/2$, supporting the microscopic origin of the stripe order of noncontractible domain walls wrapping around the torus.

\section{Additional details regarding the field theory}

\subsection{Tilted phase}
In the main text, we put forward the following Lagrangian:
\begin{equation}
    \mathcal{L} = \frac{1}{2} \bigl(\partial_{\tau}h\bigr)^2 + V\bigl[h\bigr] + \lambda \cos\bigl(2\pi h\bigr),\quad\mathrm{with}~V\bigl[h\bigr] = \frac{\rho_2}{2}\bigl(\nabla h\bigr)^2 + \frac{\rho_4}{2} \bigl(\nabla^2 h\bigr)^2 + \frac{g_4}{2}\bigl(\nabla h\bigr)^4 + \mathcal{L}_6,
    \label{eq:height_theory_SM}
\end{equation}
with $\mathcal{L}_{6} = -g_6  |\nabla h|^6 \cos(6 \varphi) - g_{12} |\nabla h|^{12} \cos(12 \varphi)$. Since we observe both vertical and horizontal stripes, we will neglect $g_6$ in the following. There are therefore 12 minima, at angles $\varphi = \mathbb{Z}2\pi/12$. We consider a solution with a uniform tilt and fluctuations around it: $h(\boldsymbol{r},\tau) = \pi^{-1}\boldsymbol{Q}\cdot\boldsymbol{r} + \delta h(\boldsymbol{r},\tau)$. First, we find $\boldsymbol{Q}$ which minimizes $V[h]$. For the sake of simplicity, we can make the assumption that $g_{12}$ has a subleading effect (except for choosing the orientation), and we can work to leading order in $g_{12}$. This leads to the simple formula,
\begin{equation}
    \frac{|\boldsymbol{Q}|}{\pi} = \sqrt{\frac{|\rho_2|}{2 g_4}}.
\end{equation}
Expanding around one of the 12 minima leads to,
\begin{equation}
    \mathcal{L} = \frac{1}{2} \bigl(\partial_{\tau}\delta h\bigr)^2 + \frac{v_{\mathrm{L}}^2}{2}\bigl(\partial_\mathrm{L} \delta h\bigr)^2 + \frac{v_{\mathrm{T}}^2}{2}\bigl(\partial_\mathrm{T} \delta h\bigr)^2,
    \label{eq:lowenergyTh_SM}
\end{equation}
with $v_\mathrm{T}^2=132\,g_{12}(|\boldsymbol{Q}|/\pi)^{10}$ and $v_\mathrm{L}^2 = 2 \rho_2$ (again, to leading order in $g_{12}$). We use indices L and T for directions longitudinal and transverse to $\boldsymbol{Q}$, respectively.

Given the numerically observed values of $|\boldsymbol{Q}|/\pi \simeq 2/5 $, $v_\mathrm{L} \simeq 1.2$, and $v_\mathrm{T}\simeq 0.3$, we extract the following field theory parameters: $g_{12}\simeq 6.5$, $\rho_2\simeq -0.72$, and $g_4\simeq 2.25$. The Goldstone action described by Eq.~\eqref{eq:lowenergyTh_SM} leads to the Matsubara Green's function $G(\boldsymbol{q},\omega)=(\omega^2 + \omega_{\boldsymbol{q}}^2)^{-1}$ for the $h$ field, with $\omega_{\boldsymbol{q}}=\sqrt{(v_\mathrm{T}q_\mathrm{T})^2 + (v_\mathrm{L}q_\mathrm{L})^2}$. Height correlation functions are then computed as,
\begin{equation}
    \begin{split}
        \Bigl\langle\delta h\bigl(\boldsymbol{r_1},\tau_1\bigr)\delta h\bigl(\boldsymbol{r_2},\tau_2\bigr)\Bigr\rangle &= \frac{1}{2\pi} \int\mathrm{d}\omega\int\mathrm{d}\boldsymbol{q}\,\mathrm{e}^{i \omega (\tau_2 - \tau_1)}\mathrm{e}^{i \boldsymbol{q}\cdot(\boldsymbol{r_2} - \boldsymbol{r_1})} \frac{1}{\omega^2 + \omega_{\boldsymbol{q}}^2} \\
        &= \frac{1}{2}\int\mathrm{d}\boldsymbol{q}\,\frac{1}{\omega_{\boldsymbol{q}}}\mathrm{e}^{i\boldsymbol{q}\cdot (\boldsymbol{r_2} - \boldsymbol{r_1})}\mathrm{e}^{-\omega_{\boldsymbol{q}} |\tau_2 - \tau_1|},
    \end{split}
\end{equation}
and $D_{ab}(\boldsymbol{q})$ of Eq.~(6) in the main text is computed by differentiation,
\begin{equation}
    D_{ab}(\boldsymbol{q})\equiv\int\mathrm{d}\boldsymbol{r_1}\int\mathrm{d}\boldsymbol{r_2}\,\mathrm{e}^{-i\boldsymbol{q}\cdot(\boldsymbol{r_2} - \boldsymbol{r_1})}\Bigl\langle\partial_a \delta h\bigl(\boldsymbol{r_1},\tau\bigr) \partial_b \delta h\bigl(\boldsymbol{r_2},\tau\bigr)\Bigr\rangle=\frac{q_aq_b}{2\omega_{\boldsymbol{q}}}.
\end{equation}

\subsection{Gauge theory dictionary}

The height theory is dual to a $U(1)$ gauge theory ~\cite{PhysRevB.21.5212,doi:10.1142/S0217984990000295,PhysRevB.69.224416,fradkin2004,Ardonne2004493,PhysRevB.65.024504,fradkin2013} for an electric field given by $\boldsymbol{E}=\boldsymbol{z}\times\nabla h$, and the tilt corresponds to a finite density of electric field lines wrapping around the torus. A dictionary between the two descriptions is given in Tab.~\ref{tab:dictionary}.

\begin{table}[!h]
    \begin{minipage}{0.8\columnwidth}
        \center
        \begin{ruledtabular}
            \begin{tabular}{cc}
                \thead{\textbf{Height theory}} & \thead{\textbf{Gauge theory}}\\
                \hline\hline\\[-0.8em]
                \makecell{Vortex of $h$ with vorticity $q$} & \makecell{Electric particle with charge $q$}\\[-0.8em]\\\hline\\[-0.8em]
                \makecell{Operator $\mathrm{e}^{i 2\pi p h}$} & \makecell{Monopole with charge $p$}\\[-0.8em]\\\hline\\[-0.8em]
                \makecell{Incommensurate stripe order with gapless phason} & \makecell{Coulomb phase with gapless photon}\\[-0.8em]\\\hline\\[-0.8em]
                \makecell{Paramagnetic phase} & \makecell{Higgs phase }\\
            \end{tabular}
        \end{ruledtabular}
        \caption{Dictionary translating between height and gauge theories.}
        \label{tab:dictionary}
    \end{minipage}
\end{table}

\subsection{Microscopic justification for the $U(1)$ gauge description of the stripe phase}
\label{ExplicitMapping}

\subsubsection{From unoriented to oriented domain walls}

Since Ising domain walls are unoriented, naively one would think they can only describe the electric field lines of a $\mathbb{Z}_2$ gauge theory: A $\mathbb{Z}_2$ electric field can only take two values, $1$ or $0$. This corresponds to having one domain wall strand, or no domain wall strand, on a given edge of the dual lattice. In the absence of electric charges, Gauss's law is ensured by the fact that domain walls are always closed.

By contrast, the electric field lines of a $U(1)$ gauge theory need to be oriented, since the electric field is now a vector $\boldsymbol{E}$. In order to have a $U(1)$ gauge description of an Ising system, it must therefore be possible to orient Ising domain walls in a local way, in order for the theory to be local. Further, if the goal is to describe a Coulomb phase, this rule for orienting domain walls should be such that the charge given by Gauss's law $Q=\nabla\cdot\boldsymbol{E}$ is dilute.

In the Higgs phases, one is dealing with a condensate of domain walls of all shapes and sizes, and it is not possible to find such a rule. Indeed, these phases are described by $\mathbb{Z}_2$, rather than $U(1)$ gauge theories~\cite{levin2012}. On the other hand, in the stripe phase, most domain walls are noncontractible, and it is possible to choose a local rule of orientation which creates an almost ``pure gauge'' configuration.

\begin{figure}[!t]
    \center
    \includegraphics[width=0.12\columnwidth,clip]{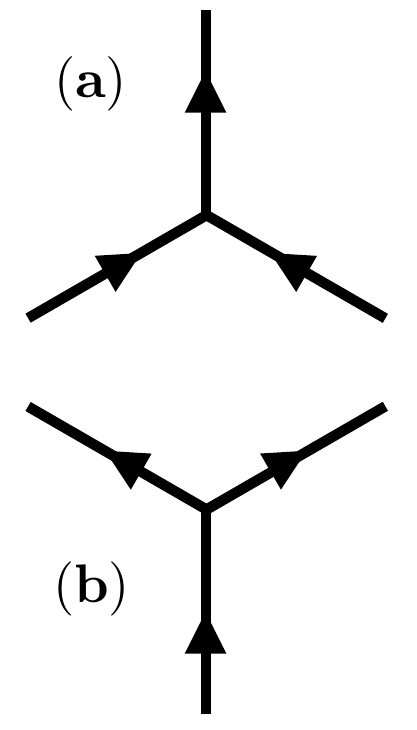}
    \caption{Edges of the honeycomb lattice are given a fixed orientation, given by a unit vector $\boldsymbol{\mathcal{E}}$. The orientation is chosen so that $\boldsymbol{\mathcal{E}}$ always has a positive vertical component. Note that these vectors are fixed, independently of the domain wall configuration.}
    \label{fig:hon_vertices_first}
\end{figure}

\begin{figure}[!t]
    \center
    \includegraphics[width=0.5\columnwidth,clip]{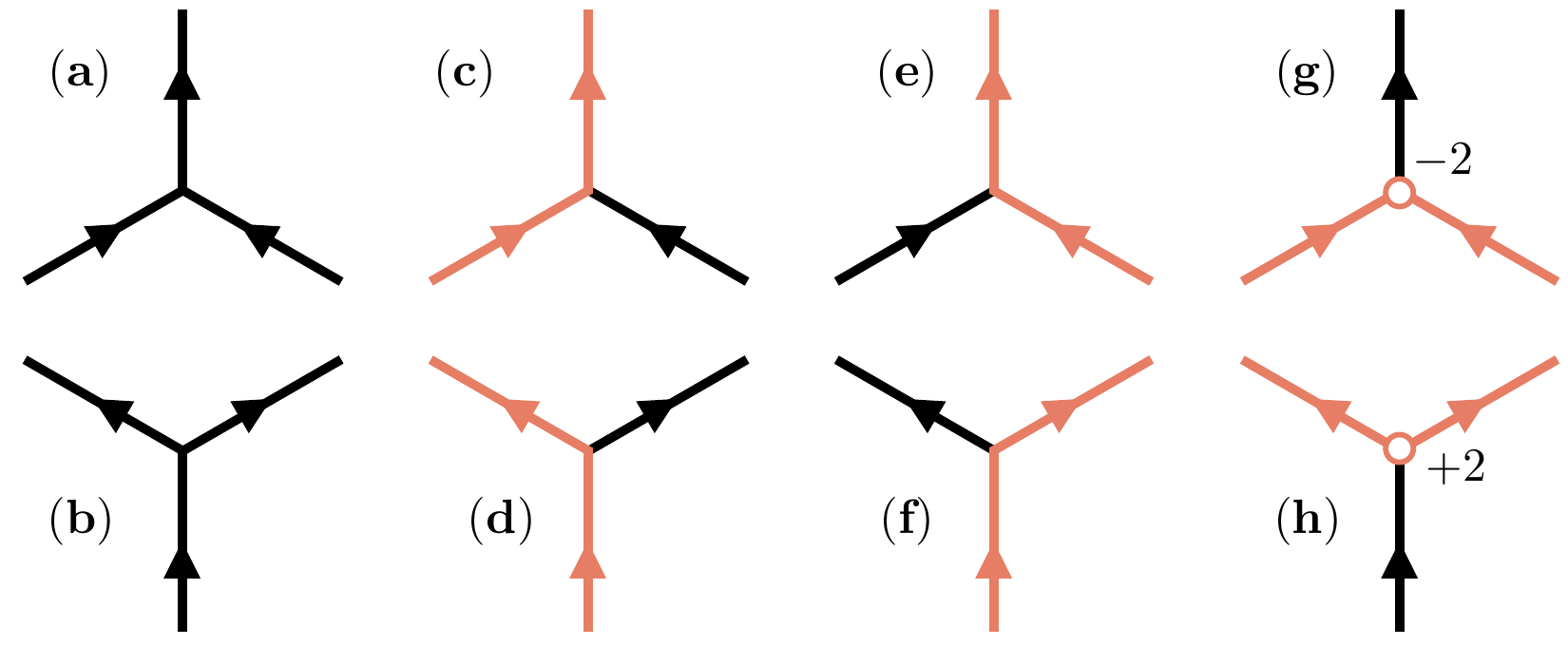}
    \caption{The arrows show the fixed unit vector $\boldsymbol{\mathcal{E}}$ on each edge.  On a given edge, the electric field is either $\boldsymbol{E}=0$ if there is no domain wall strand on it (drawn in black), or $\boldsymbol{E} = \boldsymbol{\mathcal{E}}$ if there is a domain wall strand on it (drawn in orange). (\textbf{a}) (\textbf{b}) (\textbf{c}) (\textbf{d}) (\textbf{e}) (\textbf{f}) (\textbf{g}) (\textbf{h}) Domain walls (i.e. electric field lines) are drawn in orange. Excluding configurations with dangling domain walls, there are 8 possible vertices. Gauss's law states that the charge on a given site is given by the number of outgoing electric field lines minus the number of incoming ones. In (\textbf{g}) and (\textbf{h}), the circle represents the charge, which can be either $+2$ or $-2$.}
    \label{fig:hon_vertices}
\end{figure}

For example, let us consider the case of a phase with vertical stripes. First, we give a \emph{fixed} orientation to each edge of the honeycomb lattice, see Fig.~\ref{fig:hon_vertices_first}, thereby decorating each edge with a unit vector $\boldsymbol{\mathcal{E}}$. This vector is chosen so as to always have a positive vertical component. Now, on each edge, the electric field $\boldsymbol{E}$ can only take two possible values: either $\boldsymbol{E}=0$, if there is no domain wall strand on the edge, or $\boldsymbol{E}=\boldsymbol{\mathcal{E}}$ if there is one. Finally, a charge variable is defined on each honeycomb site, and ensures Gauss's law: $Q$ counts the number of outgoing electric vectors, minus the number of incoming ones, see Fig.~\ref{fig:hon_vertices}.

\begin{figure}[!t]
    \center
    \includegraphics[width=0.9\columnwidth,clip]{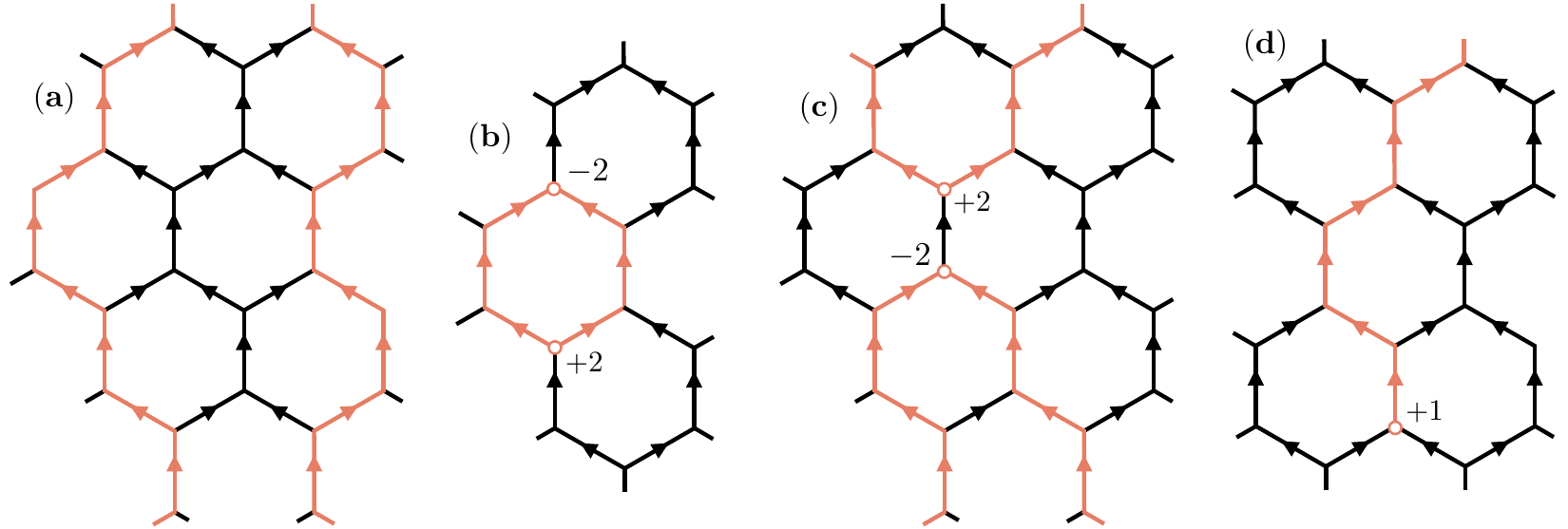}
    \caption{(\textbf{a}) As long as domain walls have a locally vertical direction, the configuration is pure gauge, i.e. there is no charge. (\textbf{b}-\textbf{c}) For $\delta\alpha\neq 0$, it becomes possible to create these configurations: either a new contractible domain wall, or recombine two noncontractible domain walls. In both cases, a vertical $\pm 2$ dipole was created. (\textbf{c}) shows a configuration after the recombination process showed in Fig.~3 of the main text. (\textbf{d}) A dangling domain wall (i.e. $\pi$ flux) has either charge $q=+1$ or $-1$.}
    \label{fig:gauge_dipole}
\end{figure}

With this definition, we can see that, as long as noncontractible domain walls go ``in the right direction'', there is no charge needed, see Fig.~\ref{fig:gauge_dipole}\,(a). This means that, deep in the stripe phase, where all domain walls are noncontractible and their local orientation fluctuates only weakly from the vertical upwards orientation, the configurations will be pure gauge, i.e., there will be almost no charge. Indeed, if a noncontractible domain wall starts changing its local orientation too much (corresponding to a overhang or hairpin turn, in the language of directed polymers~\cite{PhysRevLett.58.2087}), this would create local charges. However, deep in the stripe phase, these configurations are disfavored. This is why the $U(1)$ gauge theory description becomes warranted in the stripe phase.

On the other hand, when $\delta\alpha\neq 0$, it becomes possible to create noncontractible domain walls, which are described by an electric $q=\pm2$ dipole, see Fig.~\ref{fig:gauge_dipole}\,(b,\,c). Besides, after gauging the Ising symmetry, it becomes possible to create static excitations called $\pi$ fluxes which have a dangling domain wall. According to Gauss's law, they would have electric charge $q=\pm 1$, see Fig.~\ref{fig:gauge_dipole}\,(d).

\subsubsection{Mapping to dimer models}

\begin{figure}[!ht]
    \center
    \includegraphics[width=0.7\columnwidth,clip]{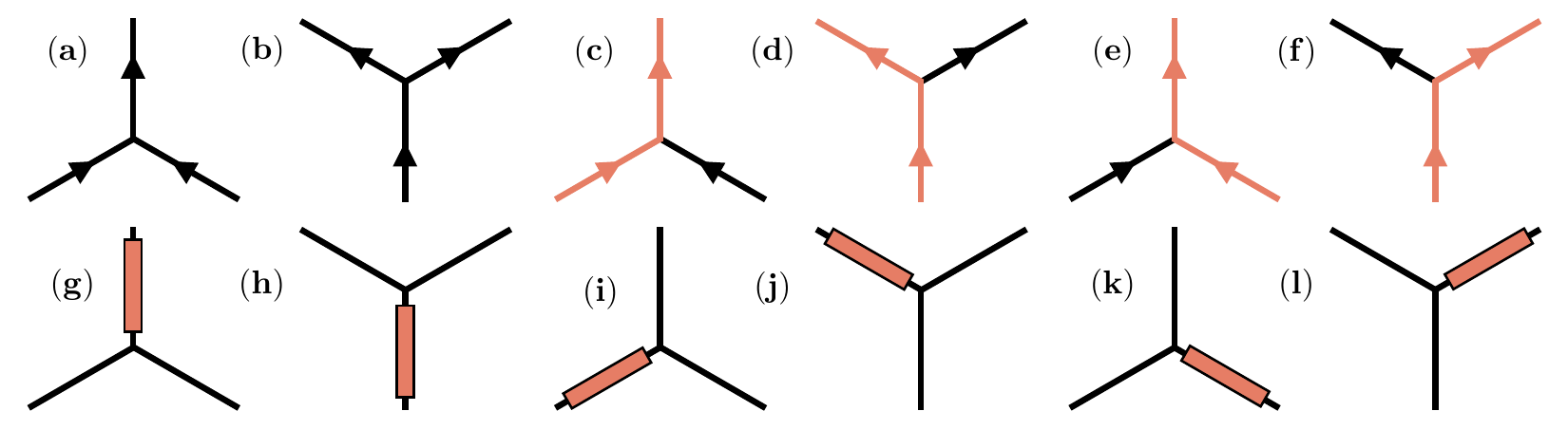}
    \caption{(\textbf{a}-\textbf{f}) There is a one-to-one mapping between pure gauge configurations of the domain walls (top row), and (\textbf{g}-\textbf{l}) dimer configurations on the honeycomb lattice (bottom row). The dimers are shown as solid rectangles.}
    \label{fig:dimer_mapping}
\end{figure}

\begin{figure}[!ht]
    \center
    \includegraphics[width=0.4\columnwidth,clip]{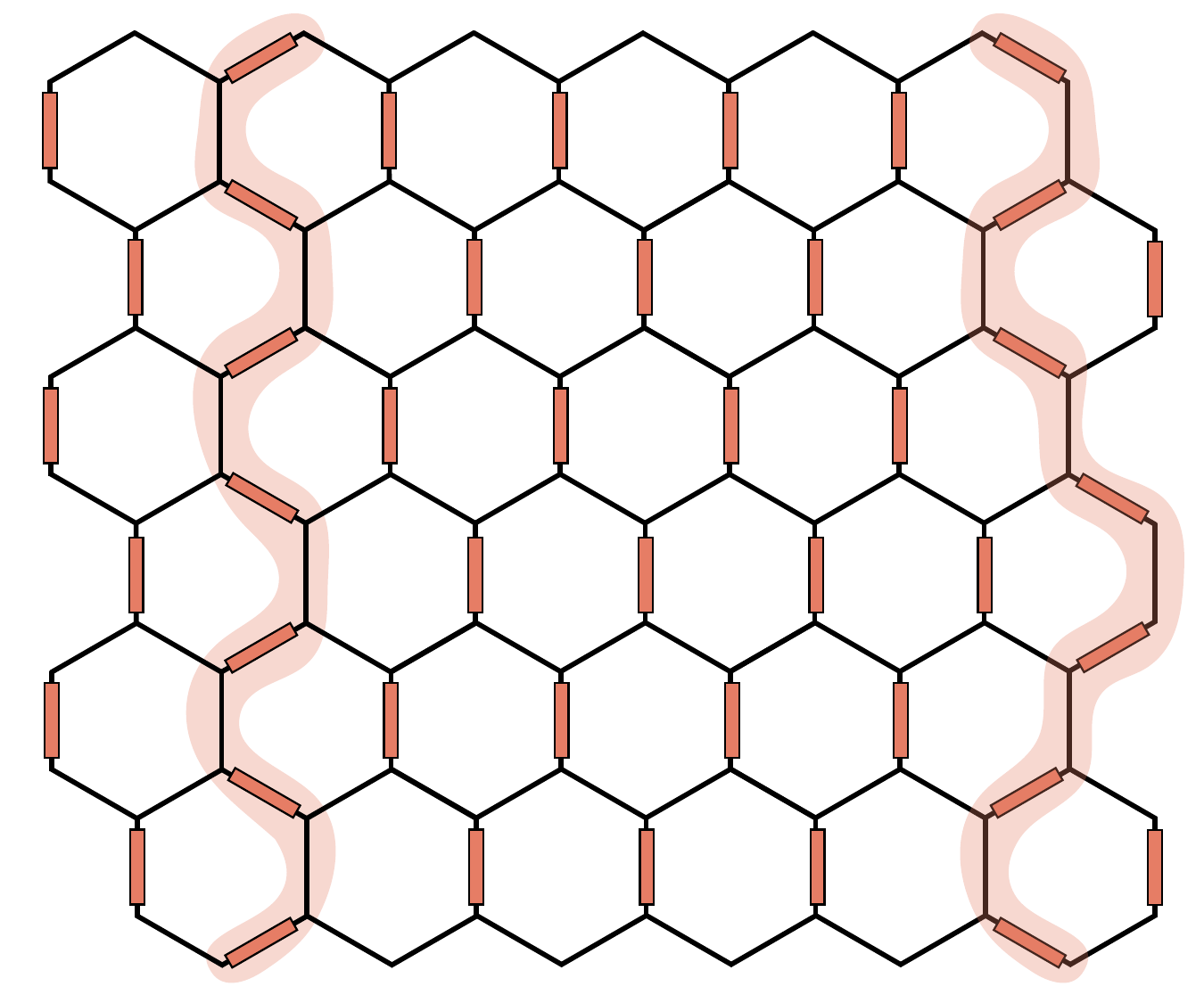}
    \caption{Mapping from a typical pure gauge configuration of domain walls (shown in translucid color) to a dimer configuration on the honeycomb lattice (shown as solid rectangles), following the one-to-one mapping rules of Fig.~\ref{fig:dimer_mapping}.}
    \label{fig:dimer_stripes}
\end{figure}

In fact, there is another way to justify a $U(1)$ gauge description of the stripe phase. As shown in Fig~\ref{fig:dimer_mapping}, there is a one-to-one mapping between ``pure-gauge'' configurations of domain walls (i.e., domain wall configurations which only include vertices (a) to (f) in Fig.~\ref{fig:dimer_mapping}) and dimer configurations on the honeycomb lattice, see panels Fig.~\ref{fig:dimer_mapping}\,(g-l). Since pure gauge configurations of domain walls dominate in the stripe phase (see Fig.~\ref{fig:dimer_stripes} for a typical configuration), it means the stripe phase is well described by a quantum dimer model on they honeycomb lattice, which has a well-known $U(1)$ gauge description. Further, the electric charges defined above can be interpreted as defects of the dimer constraint.

\end{document}